\documentclass[10pt,aps,twocolumn,superscriptaddress,showpacs,showkeys,byrevtex,bibnotes]{revtex4}
\usepackage{amssymb}
\usepackage{graphics}
\begin{document}

\hyphenation{ma-cro-state}
\hyphenation{mi-cro-state}
\hyphenation{mi-cro-ca-no-ni-cal}
\hyphenation{ca-no-ni-cal}
\preprint{Published in Physica A 335, 518-538, 2004.}
\title{Thermodynamic versus statistical nonequivalence of ensembles
for the mean-field Blume-Emery-Griffiths model}

\author{Richard S. Ellis}
\email{rsellis@math.umass.edu}
\address{Department of Mathematics and Statistics, University of
Massachusetts, Amherst, Massachusetts 01003 USA}

\author{Hugo Touchette}
\email{htouchet@alum.mit.edu}
\address{\mbox{Department of Physics and School of Computer Science,
McGill University, Montr\'eal, Qu\'ebec, Canada H3A 2A7}}

\author{Bruce Turkington}
\email{turk@math.umass.edu}
\address{Department of Mathematics and Statistics, University of
Massachusetts, Amherst, Massachusetts 01003 USA}

\begin{abstract}
We illustrate a novel characterization of nonequivalent
statistical mechanical ensembles using the mean-field Blume-Emery-Griffiths
(BEG)\ model as a test model. The novel characterization takes effect at the
level of the microcanonical and canonical equilibrium distributions of
states. For this reason it may be viewed as a \textit{statistical}
characterization of nonequivalent ensembles which extends and complements
the common \textit{thermodynamic} characterization of nonequivalent
ensembles based on nonconcave anomalies of the microcanonical entropy. By
computing numerically both the microcanonical and canonical sets of
equilibrium distributions of states of the BEG model, we show that for
values of the mean energy where the microcanonical entropy is nonconcave,
the microcanonical distributions of states are nowhere realized in the
canonical ensemble. Moreover, we show that for values of the mean energy
where the microcanonical entropy is strictly concave, the equilibrium
microcanonical distributions of states can be put in one-to-one
correspondence with equivalent canonical equilibrium distributions of
states. Our numerical computations illustrate general results relating
thermodynamic and statistical equivalence and nonequivalence of ensembles
proved by Ellis, Haven, and Turkington [\textit{J.~Stat.~Phys.~}\textbf{101}, 999 (2000)].
\end{abstract}

\maketitle

\section{Introduction}

\label{sIntro}

The microcanonical and canonical ensembles are the two main probability
distributions with respect to which the equilibrium properties of
statistical mechanical models are calculated. As is well known, the
microcanonical ensemble is a statistical mechanical expression of the
conservation of the energy for a closed or isolated system whereas the
canonical ensemble models a system in thermal interaction with a heat
reservoir having a constant temperature \cite{reif1965,pathria1996}.
Although the two ensembles model two different physical situations, it is
widely assumed that the ensembles give equivalent results in the
thermodynamic limit; i.e., in the limit in which the volume of the system
tends to infinity. The typical argument used to motivate this equivalence is
given in the classic text of Landau and Lifshitz \cite{landau1991}. Although
``the canonical distribution is `spread' over a certain range of energies,''
``the width of this range \ldots is negligible for a macroscopic body.'' The
conclusion is that, in the thermodynamic limit, the canonical ensemble can
be considered to be an ensemble of fixed mean energy or, in other words, a
microcanonical ensemble. The texts \cite
{gibbs1902,reif1965,wannier1966,huang1987,balian1991} contain similar
arguments.

In order to substantiate the argument of Landau and Lifshitz, it must be
proved that the energy per particle converges to a constant in the
infinite-volume limit of the canonical ensemble. This convergence can be
proved to hold for noninteracting systems such as the perfect gas and for a
variety of weakly interacting systems. For general systems, however, neither
is this convergence valid nor is the conclusion true concerning ensemble
equivalence which this convergence is intended to motivate. In fact, in the
past three and a half decades, numerous statistical models have been
discovered having microcanonical equilibrium properties that cannot be
accounted for within the framework of the canonical ensemble. This lack of
correspondence between the microcanonical and canonical ensembles has
profound consequences for it implies that one is forbidden to substitute the
mean-energy variable for the temperature variable, and vice versa, when
parameterizing the equilibrium properties of systems. In such cases of
nonequivalence, the questions of determining which of the two ensembles is
the more fundamental and which ensemble is the one realized in the
laboratory are of fundamental interest.

Until now, the phenomenon of nonequivalent ensembles has been identified and
analyzed almost exclusively by determining regions of the mean energy where
the microcanonical entropy function is anomalously nonconcave or by
determining regions of the mean energy where the heat capacity, calculated
microcanonically, is negative. This thermodynamic approach to the problem of
nonequivalent ensembles has been propounded by a number of people, including
Lynden-Bell and Wood \cite{lynden1968}, who in the 1960's were among the
first to observe negative heat capacities in certain gravitational many-body
systems (see also \cite
{thirring1970,hertel1971,thirring1983,lynden1999,padmanabhan1990,gross2001}). More
recently, nonconcave anomalies in the microcanonical entropy as well as
negative heat capacities have been observed in models of fluid turbulence
\cite{gaglioti1995,robert1991,kiessling1997,ellis2002} and models of plasmas
\cite{smith1990}, in addition to long-range and mean-field spin models,
including the mean-field XY model \cite{dauxois2000} and the mean-field
Blume-Emery-Griffiths (BEG) model \cite{barre2001,barre2002}.

The existence of such nonconcave anomalies invalidates yet another tacit
principle of statistical mechanics which states that the one should always
be able to express the microcanonical entropy, the basic thermodynamic
function for the microcanonical ensemble, as the Legendre-Fenchel transform
of the free energy, the basic thermodynamic function for the canonical
ensemble. Indeed, if the microcanonical entropy is to be expressed as the
Legendre-Fenchel transform of the canonical free energy, then the former
function must necessarily be concave on its domain of definition. Hence, if
the microcanonical entropy has nonconcave regions, then expressing it as a
Legendre-Fenchel transform is impossible. When this occurs, we say that
there is thermodynamic nonequivalence of ensembles.

For the BEG model, the existence of nonconcave anomalies in the
microcanonical entropy was shown by Barr\'{e}, Mukamel and Ruffo via a
Landau expansion of this quantity \cite{barre2001}. The aim of the present
paper is to extend their study of nonequivalent ensembles by showing how the
thermodynamic nonequivalence of the microcanonical and canonical ensembles
for this model reflects a deeper level of nonequivalence that takes place at
the statistical level of its equilibrium distribution of states. We carry
this out via numerical calculations both at the thermodynamic level and at
the statistical level, illustrating, in particular, a striking statistical
consequence of the nonconcavity of the microcanonical entropy. Namely, we
demonstrate that if the microcanonical entropy is nonconcave at a
mean-energy value $u$---i.e., if thermodynamic nonequivalence of ensembles
holds---then nonequivalence of ensembles holds at the statistical level in
the following sense: none of the equilibrium statistical distributions of
states calculated in the microcanonical ensemble at that value of $u$ can be
realized in the canonical ensemble. Furthermore, we demonstrate that if the
microcanonical entropy is strictly concave at $u$---i.e., if thermodynamic
equivalence of ensembles holds---then all the statistical distributions of
states found in the microcanonical ensemble at that value of $u$ can be put
in one-to-one correspondence with equivalent distributions of the canonical
ensemble.

Our numerical findings illustrate clearly and directly a number of results
on equivalence and nonequivalence of ensembles that are valid for a wide
class of statistical mechanical systems. These results, derived recently by
Ellis, Haven and Turkington, establish different levels of correspondence
between microcanonical equilibrium macrostates and canonical equilibrium
macrostates in terms of concavity properties of the microcanonical entropy.
These results, as well as an overview of applications to turbulence, can be
found in their comprehensive paper \cite{ellis2000}. A specific application
to a model of geophysical fluid turbulence is treated in their paper 
\cite{ellis2002}.

The organization of the present paper is as follows. In Section \ref{sThermo}
we introduce the basic thermodynamic functions in the two ensembles, the
microcanonical entropy and the canonical free energy. We also motivate the
definitions of the sets of equilibrium statistical distributions of states
for the two ensembles in the thermodynamic limit. This motivation is based
in part on the theory of large deviations, a branch of probability that
studies the exponential decay of probabilities and is perfectly suited for
analyzing asymptotic properties of the two ensembles. In Section
\ref{sEquivalence} we discuss the equivalence and nonequivalence of ensembles
first at the thermodynamic level and then at the statistical level. In
Section \ref{sBEG}, we finally present our numerical calculations that
illustrate thermodynamic and statistical equivalence and nonequivalence of
ensembles for the BEG model. Because our goal is to emphasize the physical
ideas, we have omitted almost all mathematical details and have occasionally
compromised perfect mathematical accuracy when it benefits the exposition.
For complete mathematical details concerning equivalence and nonequivalence
of ensembles and the theory of large deviations, the reader is referred to
\cite{ellis2000,ellis1985,dembo1998}.

\section{Thermodynamic ensembles and large deviations principles}

\label{sThermo}

\subsection{Microcanonical ensemble}

\label{sMicro} The main quantity characterizing the thermodynamic properties
of statistical mechanical systems in the microcanonical ensemble is the
entropy function. This quantity is defined in terms of the probability
measure of all microstates of a system having the same value of the mean
energy \cite{reif1965,pathria1996}. To be precise, suppose that the system
in question is composed of $n$ particles, and denote the microstates of that
system by the joint state $x^n=(x_1,x_2,\ldots ,x_n)$, where $x_i$
represents the state of the $i^{\rm th}$ particle taking values in some
state space $\mathcal{X}$. The set of all microstates is the $n$-fold
product $\mathcal{X}^n$. Moreover, let us denote by $U(x^n)$ the energy or
Hamiltonian of the system as a function of the microstates $x^n$ and by
$u(x^n)$ the mean energy or energy per particle, defined as 
\begin{equation}
u(x^n)=\frac{U(x^n)}n.
\end{equation}

In terms of this notation the microcanonical entropy function $s(u)$ is
defined as 
\begin{equation}
s(u)=\lim_{n\rightarrow \infty }\frac 1n\ln P\{u(x^n)\in du\},  \label{s1}
\end{equation}
where
\begin{equation}
P\{u(x^n)\in du\}=\int_{\{x^n:u(x^n)\in du\}}P(dx^n)  \label{s11}
\end{equation}
is the probability measure of all microstates $x^n$ lying in the
infinitesimal mean-energy ball $du$ centered at $u$ \cite{ellis2000}. The
probability measure $P(dx^n)$ in (\ref{s11}) is the \textit{a priori measure}
on $\mathcal{X}^n$, which is taken to be the uniform measure in accordance
with Boltzmann's equiprobability hypothesis \cite{reif1965,pathria1996}. In
order for this probability measure to be well defined, the configuration
space $\mathcal{X}^n$ is assumed to be bounded. As we will soon explain,
conditioning $P$ on the set of configurations $\{x^n:u(x^n)\in du\}$ defines
the microcanonical ensemble. We call this set the \textit{microcanonical
conditioning set}.

We find it convenient to re-express the definition of $s(u)$ in (\ref{s1})
via the formula
\begin{equation}
P\{u(x^n)\in du\}\asymp e^{ns(u)}.  \label{ldp1}
\end{equation}
This is done in order to emphasize the facts that $P\{u(x^n)\in du\}$ has,
to a first degree of approximation, the form of an exponential that decays
with $n$ and that the exponential decay rate is the microcanonical entropy.
The heuristic sign ``$\asymp $'' is used here instead of the approximation
sign ``$\approx $'' in order to stress that the dominant term describing the
asymptotic behavior of $P\{u(x^n)\in du\}$, as $n\rightarrow \infty $, is
the exponential function $e^{ns(u)}$.

In the theory of large deviations, the exponential asymptotic property of
thermodynamic probability measures is referred to as a \textit{large
deviation principle} (LDP). Detailed explanations of this theory are
available in a number of references including
\cite{dembo1998,ellis1985,ellis1995,ellis1999,oono1989}. For our purpose it is
important to note that an LDP can be formulated not only for the energy per
particle, but also for more general macroscopic variables such as the total
spin per particle or the vector of occupation numbers. As we point out in
the next paragraph, the LDP is the basic tool for deriving the well known
variational maximum-entropy principle characterizing the equilibrium
macrostates of the associated macroscopic variables in the microcanonical
ensemble.

The macroscopic variables that we consider are quantities $L(x^n)$ that take
values in a space of macrostates $\mathcal{L}$, and that are assumed to
satisfy an LDP expressed by
\begin{eqnarray}
P\{L(x^n)
\in
dL\} &=&\int_{\{x^n:L(x^n)\in dL\}}P(dx^n)  \label{ldp2} \\
&\asymp &e^{ns(L)}.  \nonumber
\end{eqnarray}
In this formula $L$ is an element of $\mathcal{L}$, and $P\{L(x^n)\in dL\}$
represents the probability measure of all microstates $x^n$ such that $
L(x^n) $ lies in the infinitesimal ball $dL$ with center $L$. In addition,
we assume that the energy per particle $u(x^n)$ can be written
asymptotically as a function of $L(x^n)$ in the following sense. There
exists a bounded, continuous function $u(L)$, called the \textit{energy
representation function}, with the property that $u(x^n)=u(L(x^n))$ for all
microstates $x^n$ or, more generally, that
\begin{equation}
|u(L(x^n))-u(x^n)|\rightarrow 0  \label{ulxn}
\end{equation}
uniformly over all microstates $x^n$ as $n\rightarrow \infty $. In Section
\ref{sBEG} we give a concrete example, for the BEG model, of a macroscopic
variable $L(x^n)$ for which assumptions (\ref{ldp2}) and (\ref{ulxn}) are
valid. In the case of the BEG model, the energy representation will in fact
be seen to be exact in the sense that $u(x^n)=u(L(x^n))$ for all $x^n$ and $n$.
However, the properties discussed in this paper are valid also for models for
which the limit (\ref{ulxn}) holds.

Under assumptions (\ref{ldp2}) and (\ref{ulxn}) on the macroscopic variable
$L(x^n)$, it is easily seen that the most probable macrostates $L$ for
configurations lying in the microcanonical conditioning set $\{x^n:u(x^n)\in
du\}$ are those that maximize the entropy function $s(L)$ subject to the
energy constraint $u(L)=u$ \cite{ellis2000}. These constrained maximizers of
$s(L)$ compose the set $\mathcal{E}^u$ of \textit{microcanonical equilibrium
macrostates} associated with a given mean-energy value $u$; in symbols,
\begin{equation}
\mathcal{E}^u=\{L:L\textrm{ maximizes }s(L)\textrm{ with }u(L)=u\}.
\label{micro1}
\end{equation}

\begin{widetext}
The physical importance of equilibrium macrostates in statistical mechanics
stems from the fact that any macrostate not in $\mathcal{E}^u$ has an
exponentially small probability of being observed given that the energy per
particle of the system is fixed at the value $u$. This can be seen by
introducing the microcanonical probability measure $P^u$, which is defined
by conditioning the a priori measure $P$ on the microcanonical conditioning
set \cite{ellis2000}; in symbols, $P^u(dx^n) = P\{dx^n \, | \, u(x^n) \in
du\}$. Thus for $L \in \mathcal{L}$ we have
\begin{eqnarray}
P^u\{L(x^n)
\in
dL\} &=&P\{L(x^n)
\in
dL \, | \, u(x^n)\in du\} \\
&=&\frac{P\{\{x^n:L(x^n)\in dL\}\cap \{x^n:u(x^n)\in du\}\}}{P\{u(x^n)\in
du\}}.  \nonumber
\end{eqnarray}
\end{widetext}
As shown in \cite{ellis2000}, the LDP's (\ref{ldp1}) and (\ref{ldp2}) and
the asymptotic relationship (\ref{ulxn}) yield the LDP
\begin{equation}
P^u\{L(x^n)\in dL\}\asymp e^{-nI^u(L)},
\end{equation}
where
\begin{equation}
I^u(L)=\left\{
\begin{array}{lll}
s(u)-s(L) &  & \textrm{if }u(L)=u \\
\infty &  & \textrm{otherwise.}
\end{array}
\right.  \label{mrate}
\end{equation}

By the general theory of large deviations, $I^u(L)$ is nonnegative for any
macrostate $L\in \mathcal{L}$. Hence if $I^u(L)>0$, then the microcanonical
probability that $L(x^n)$ is near $L$ goes to $0$, as $n\rightarrow \infty$,
at the exponential decay rate $I^u(L)$. This observation motivates our
definition of the set $\mathcal{E}^u$ of microcanonical equilibrium
macrostates to be the set of macrostates $L$ for which $I^u(L)$ attains its
minimum of $0$, and thus for which $s(L)$ is maximized under the constraint
$u(L)=u$. As a consequence of this definition, we have the following
variational formula for the microcanonical entropy:
\begin{equation}
s(u)=\sup_{\{L:u(L)=u\}}s(L)=s(L^u),  \label{ment1}
\end{equation}
where $L^u$ is any macrostate contained in $\mathcal{E}^u$.\footnote{
Since $s(L)$ is nonpositive, $s(u)$ is also nonpositive. However, $s(u)$ can
equal $-\infty $. Throughout this paper, we shall restrict ourselves to
values of $u$ for which $s(u)>-\infty $. \label{foot1}}

\subsection{Canonical ensemble}

\label{sCanon}

While the microcanonical ensemble is defined in terms of a fixed value of
the mean energy $u$, the canonical ensemble is defined in terms of a fixed
value of the inverse temperature $\beta $. In the canonical ensemble the
relevant probability measure on $\mathcal{X}^n$ is the Gibbs measure
\begin{equation}
P_\beta (dx^n)=\frac 1{Z_n(\beta )}e^{-\beta nu(x^n)}P(dx^n).
\end{equation}
In this formula, $Z_n(\beta )$ is the $n$-particle partition function
defined by
\begin{equation}
Z_n(\beta )=\int_{\mathcal{X}^n}e^{-\beta nu(x^n)}P(dx^n),  \label{znbeta}
\end{equation}
in terms of which we define the \textit{canonical free energy}
\begin{equation}
\varphi (\beta )=-\lim_{n\rightarrow \infty }\frac 1n\ln Z_n(\beta ).
\label{fren1}
\end{equation}
This last quantity plays an analogous role in the asymptotic analysis of the
canonical ensemble as the microcanonical entropy plays in the asymptotic
analysis of the microcanonical ensemble.

\begin{widetext}
As we did in the case of the microcanonical ensemble, we now state the LDP
for $L(x^n)$ with respect to the canonical ensemble and then use this
principle to define the set of canonical equilibrium macrostates. For any
macrostate $L$ and any microstates $x^n$ satisfying $L(x^n)\in dL$, the
continuity of the energy representation function implies that $u(x^n)$ is
close to $u(L)$. Hence, we expect that
\begin{eqnarray}
P_\beta \{L(x^n)
\in
dL\} &=&\int_{\{x^n:L(x^n)\in dL\}}P_\beta (dx^n) \\
&=&\frac 1{Z_n(\beta )}\int_{\{x^n:L(x^n)\in dL\}}e^{-\beta nu(x^n)}\,P(dx^n)
\nonumber
\end{eqnarray}
is close to
\begin{equation}
\frac 1{Z_n(\beta )}\,e^{-\beta nu(L)}\,\int_{\{x^n:L(x^n)\in
dL\}}P(dx^n)=\frac 1{Z_n(\beta )}\,e^{-\beta nu(L)}\,P\{x^n:L(x^n)\in dL\}.
\end{equation}
\end{widetext}
Substituting into this formula the large deviation estimate (\ref{ldp2}) for
$P\{L(x^n)\in dL\}$ and the limit (\ref{fren1}) relating $Z_n(\beta )$ and
$\varphi (\beta )$, we obtain the LDP
\begin{equation}
P_\beta \{L(x^n)\in dL\}\asymp e^{-nI_\beta (L)},  \label{ldp3}
\end{equation}
where
\begin{equation}
I_\beta (L)=\beta u(L)-s(L)-\varphi (\beta ).
\end{equation}

The function $I_\beta (L)$ is nonnegative for any macrostate $L$. As in the
case of the microcanonical ensemble, for any macrostate $L$ satisfying
$I_\beta (L)>0$ the LDP (\ref{ldp3}) shows that the corresponding canonical
probability $P_\beta \{L(x^n)\in dL\}$ converges to 0 exponentially fast. As
a result, such macrostates are not observed in the thermodynamic limit. The
conclusion that we draw from the LDP (\ref{ldp3}) is that, with respect to
the Gibbsian probability measure $P_\beta $, the most probable macrostates
are those for which $I_\beta (L)$ attains its minimum of 0 or, equivalently,
those for which the quantity $\beta u(L)-s(L)$ is minimized for a fixed
value of $\beta $. Accordingly, the set $\mathcal{E}_\beta $ of \textit{
canonical equilibrium macrostates} associated with a fixed value of $\beta $
is defined by
\begin{equation}
\mathcal{E}_\beta =\{L:\beta u(L)-s(L)\textrm{ is minimized}\}.  \label{can1}
\end{equation}

We end this section by motivating the existence of the limit (\ref{fren1})
that defines the canonical free energy $\varphi (\beta )$. In the definition
(\ref{znbeta}) of $Z_n(\beta )$, we first use (\ref{ulxn}) to replace the
energy per particle $u(x^n)$ by $u(L(x^n))$. Since the macroscopic variable
$L(x^n)$ takes values in $\mathcal{L}$, we can rewrite the resulting
expression for $Z_n(\beta )$ not as an integral over the set $\mathcal{X}^n$
of microstates, but as an integral over $\mathcal{L}$. Applying the LDP
(\ref{ldp2}) for $P(L(x^n)\in dL)$, we obtain
\begin{eqnarray}
Z_n(\beta ) &=&\int_{\mathcal{L}}e^{-\beta nu(L)}P\{L(x^n)\in dL\} \\
&\asymp &\int_{\mathcal{L}}e^{-n[\beta u(L)-s(L)]}dL  \nonumber \\
&\asymp &\exp \left( -n\inf_{L\in \mathcal{L}}\{\beta u(L)-s(L)\}\right) .
\nonumber
\end{eqnarray}
The last step is a consequence of Laplace's method, which states that, as
$n\rightarrow \infty $, the main exponential contribution to the integral
comes from the largest value of the integrand or equivalently the smallest
exponent \cite{ellis1985,ellis1995}. These calculations motivate the
variational formula
\begin{eqnarray}
\varphi (\beta ) &=&\inf_{L\in \mathcal{L}}\{\beta u(L)-s(L)\}  \label{lf3}
\\
&=&\beta u(L_\beta )-s(L_\beta ),  \nonumber
\end{eqnarray}
where $L_\beta $ is any member of $\mathcal{E}_\beta $. We call this the
\textit{macrostate representation} of $\varphi (\beta )$. Formula (\ref{lf3})
can be derived rigorously from the LDP (\ref{ldp2}) using Varadhan's
theorem \cite{ellis2000}.

We denote by $\mathcal{U}$ the set of mean energy values $u$ for which $s(u)
> -\infty$. Any $u \in \mathcal{U}$ is called an
\textit{admissible value} of the mean energy. Using a similar chain of arguments as
in the preceding paragraph, one can also write
\begin{eqnarray}
Z_n(\beta ) &=&\int_{\mathcal{U}}e^{-\beta nu}P\{u(x^n)\in du\} \\
&\asymp &\int_{\mathcal{U}}e^{-n[\beta n-s(u)]}du  \nonumber \\
&\asymp &\exp \left( -n\inf_{u\in \mathcal{U}}\{\beta u-s(u)\}\right).
\nonumber
\end{eqnarray}
This asymptotic formula motivates the fundamental relationship
\begin{equation}
\varphi (\beta )=\inf_{u\in \mathcal{U}}\{\beta u-s(u)\},  \label{lf1}
\end{equation}
which expresses the thermodynamic free energy $\varphi (\beta )$ as the
\textit{Legendre-Fenchel transform} of the microcanonical entropy $s(u)$
\cite{ellis2000}.\ We call this formula the \textit{thermodynamic
representation} of $\varphi (\beta )$.

The same formula for $\varphi (\beta )$ can also be derived by rewriting the
infimum over $L$ in (\ref{lf3}) as an infimum over all mean-energy values $u$
followed by a constrained infimum over all $L$ satisfying $u(L)=u$. Using
(\ref{ment1}), we obtain 
\begin{eqnarray}
\varphi (\beta ) &=&\inf_{u\in \mathcal{U}}\,\inf_{\{L:u(L)=u\}}\{\beta
u(L)-s(L)\} \\
&=&\inf_{u\in \mathcal{U}}\left\{ \beta u-\inf_{\{L:u(L)=u\}}\{s(L)\}\right\}
\nonumber \\
&=&\inf_{u\in \mathcal{U}}\{\beta u-s(u)\}.  \nonumber
\end{eqnarray}
In the case where $s(u)$ is a strictly concave differentiable function of $u$
and $\beta $ is in the range of $s^{\prime }$, this Legendre-Fenchel
transform reduces to the usual differential form of the Legendre transform.
This is given by
\begin{equation}
\varphi (\beta )=\beta u(\beta )-s(u(\beta )),  \label{lf2}
\end{equation}
where $u(\beta )$ is the unique solution of the equation $\beta =s^{\prime}(u)$.

\section{Equivalence and nonequivalence of ensembles}

\label{sEquivalence}

\subsection{Thermodynamic level}

\label{sThermolevel}

The problem of the equivalence or nonequivalence of the microcanonical and
canonical ensembles at the thermodynamic level is fundamentally related to
the properties of the Legendre-Fenchel transform, and especially its
invertibility or noninvertibility properties as a functional transform.
Equation (\ref{lf1}) expresses $\varphi (\beta )$ as the Legendre-Fenchel
transform of $s(u)$. A basic question is whether one can invert this
Legendre-Fenchel transform by applying the same transform to $\varphi (\beta
)$ so as to obtain 
\begin{equation}
s(u)=\inf_\beta \{\beta u-\varphi (\beta )\}.
\end{equation}

\begin{figure*}
\resizebox{0.75\textwidth}{!}{\includegraphics{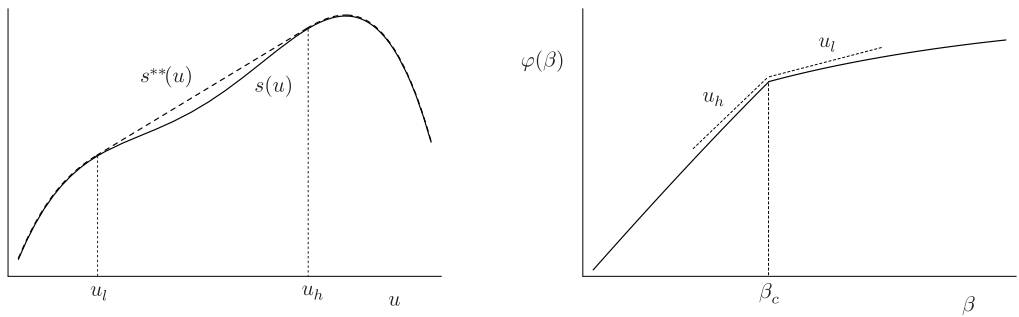}}
\caption{(Left) Plot of a nonconcave entropy function $s(u)$ together with its
concave envelope $s^{**}(u)$. The nonconcavity region equals the
open interval $(u_l,u_h)$. (Right) The corresponding free energy function 
$\varphi({\protect\beta})$ obtained by calculating the Legendre-Fenchel
transform of $s(u)$. 
The region of nonconcavity of $s(u)$ is signalled
by the appearance of a point
$\protect{\beta_c}$ where $\varphi({\protect\beta})$ is nondifferentiable.
The quantity $\protect{\beta_c}$ equals
the slope of the affine part of $s^{**}(u)$, and the left-hand 
and right-hand derivatives of $\varphi$ at $\protect{\beta_c}$ 
equal $u_h$ and $u_l$, respectively.}
\label{fe3}
\end{figure*}

We claim that such an inversion of the Legendre-Fenchel transform is valid
if and only if $s(u)$ is concave on its domain of definition. The
sufficiency can be seen by introducing the function 
\begin{equation}
s^{**}(u)=\inf_\beta \{\beta u-\varphi (\beta )\},  \label{s**}
\end{equation}
which is concave on its domain of definition and equals the minimal concave
function majorizing $s(u)$ for all $u$ \cite{ellis2000,ellis1985}. We call
$s^{**}(u)$ the \textit{concave hull} of $s(u)$ and depict it in Figure
\ref{fe3}. It follows that if $s(u)$ is concave on its domain of definition,
then $s$ and $s^{**}$ coincide. Replacing $s^{**}(u)$ by $s(u)$ in (\ref{s**})
expresses the microcanonical entropy $s(u)$ as the Legendre-Fenchel
transform of the free energy $\varphi (\beta )$. Conversely, since any
function written as the Legendre-Fenchel transform of another function is
automatically concave on its domain of definition, it follows that if $s(u)$
is not concave on its domain of definition, then it cannot be written as the
Legendre-Fenchel transform of the free energy.

This discussion motivates the following definitions. We define the two
ensembles to be \textit{thermodynamically equivalent} at $u$ if
$s(u)=s^{**}(u)$; in this case we say that $s$ is concave at $u$. In the
opposite case---namely, if $s(u)\neq s^{**}(u)$---we call the two ensembles 
\textit{thermodynamically nonequivalent} at $u$.\footnote{
The occurrence of a negative heat capacity in the microcanonical ensemble
provides a sufficient but not necessary criterion for characterizing the
microcanonical and canonical ensembles as being thermodynamically
nonequivalent. Indeed, the microcanonical heat capacity calculated at $u$
can be positive even if $s(u)\neq s^{**}(u)$. On the other hand, a negative
microcanonical heat capacity at $u$ implies necessarily that $s(u)\neq
s^{**}(u)$.} If $s(u)\neq s^{**}(u)$, then $s$ is said to be nonconcave at
$u$. For later reference we say that $s$ is strictly concave at $u$ if
$s(u)=s^{**}(u)$ and $s^{**}$ is strictly concave at $u$ in the sense that the
graph of $s^{**}$ is not flat around u.

The noninvertibility of the Legendre-Fenchel transform for nonconcave
functions has no effect on how $\varphi $ is to be calculated from $s$. The
canonical free energy $\varphi(\beta)$ is always a concave function of the
inverse temperature $\beta$; regardless of the form of $s(u)$,
$\varphi(\beta)$ can always be expressed via the fundamental Legendre-Fenchel
relationship (\ref{lf1}). As we have just seen, however, $s(u)$ can be
expressed as the Legendre-Fenchel of $\varphi(\beta)$ if and only if $s(u)$
is concave on its domain of definition. In this sense, the microcanonical
ensemble is more fundamental than the canonical ensemble.

This apparent superiority of the microcanonical ensemble over the canonical
ensemble does not prevent us from deriving a criterion based entirely on the
canonical ensemble for verifying that the two ensembles are
thermodynamically equivalent. Indeed, suppose that $\varphi (\beta )$ is
differentiable for all $\beta $. Then the G\"{a}rtner-Ellis Theorem
guarantees that, with respect to the a priori measure $P$, the energy per
particle $u(x^n)$ satisfies the LDP with entropy function $s(u)$ given by
the Legendre-Fenchel transform of $\varphi (\beta )$ \cite
{dembo1998,ellis1985}. Because $\varphi (\beta )$ is assumed to be
everywhere differentiable, the general theory of these transforms guarantees
that $s(u)$ is strictly concave on its domain of definition. We conclude
that if $\varphi (\beta )$ is everywhere differentiable, then thermodynamic
equivalence of ensembles holds for all admissible values of the mean
energy.\footnote{
The differentiability of $\varphi ({\beta })$ for all $\beta $ also implies
that with respect to the canonical ensemble the energy per particle $u(x^n)$
converges to the constant value $\varphi ^{\prime }({\beta })$ in the limit
$n\rightarrow \infty $ \cite{ellis1985}.\label{uxn}} This can be expressed in
more physical terms by saying that the absence of a first-order phase
transition in the canonical ensemble implies that the ensembles are
equivalent at the thermodynamic level. Unfortunately, the converse statement
does not hold as the nondifferentiability of $\varphi $ at some $\beta $
corresponds to one of the following: either $s(u)$ is not concave over some
range of mean-energy values or $s(u)$ is concave, but not strictly concave,
over some range of mean-energy values (see Figure \ref{fe3}).

In the next subsection we examine the equivalence and nonequivalence of
ensembles at a higher level; namely, that of equilibrium macrostates. Among
other results we will see that the strict concavity of $s$ at some
mean-energy value $u$ implies a strong form of equivalence that we call full
equivalence.

\subsection{Macrostate level}

\label{sMacrolevel}

At the level of equilibrium macrostates, the natural questions to consider
for characterizing the equivalence or nonequivalence of the microcanonical
and canonical ensembles are the following.\ For every $\beta $ and every
$L_\beta $ in the set $\mathcal{E}_\beta $ of canonical equilibrium
macrostates, does there exist a value of $u$ such that $L_\beta $ lies in
the set $\mathcal{E}^u$ of microcanonical equilibrium macrostates?
Conversely, for every $u$ and every $L^u\in \mathcal{E}^u$, does there exist
a value of $\beta $ such that $L^u\in \mathcal{E}_\beta $? In trying to
relate the macrostate level of equivalence and nonequivalence of ensembles
with the thermodynamic level of equivalence and nonequivalence, we also ask
the following. Are there thermodynamic conditions expressed in terms of
properties of $s(u) $ or $\varphi (\beta )$ ensuring a correspondence or a
lack of correspondence between the members of $\mathcal{E}^u$ and those of
$\mathcal{E}_\beta $? In particular, does equivalence of ensembles at the
thermodynamic level implies equivalence of ensembles at the level of
equilibrium macrostates?

In \cite{ellis2000} Ellis, Haven, and Turkington have provided precise
answers to these questions expressed in terms of relationships between the
solutions of the constrained minimization problem (\ref{micro1}) that
characterizes the set $\mathcal{E}^u$ of microcanonical equilibrium
macrostates and the solutions of the dual unconstrained minimization problem
(\ref{can1}) that characterizes the set $\mathcal{E}_\beta $ of canonical
equilibrium macrostates. Their main results are summarized in items 1-4
below; they apply in general to any statistical mechanical model that have
macroscopic variables $L(x^n)$ satisfying an LDP as in (\ref{ldp2}) with an
entropy function $s(L)$ and that have an energy representation function
$u(L) $ satisfying (\ref{ulxn}). In order to simplify the presentation, we
consider only mean-energy values $u$ lying in the interior of the domain of
definition of $s$, and we assume that $s$ is differentiable at all such $u$.
The reader is referred to \cite{ellis2000} for complete proofs of more
general results that hold under weaker assumptions.

\begin{enumerate}
\item  \textit{Canonical equilibrium macrostates can always be realized
microcanonically}. Let $\beta $ be given. Then 
\begin{equation}
\mathcal{E}_\beta =\bigcup_{u\in u(\mathcal{E}_\beta )}\mathcal{E}^u,
\label{canmicro}
\end{equation}
where $u(\mathcal{E}_\beta )$ denotes the set of mean-energy values $u$ that
can be written as $u(L)$ for some $L\in \mathcal{E}_\beta $. Formula (\ref
{canmicro}) implies that for any macrostate $L_\beta \in \mathcal{E}_\beta $
there exists $u\in u(\mathcal{E}_\beta )$ such that $L_\beta \in \mathcal{E}
^u$.

\item  \textit{Full equivalence}. If $s$ is strictly concave at $u$, then
$\mathcal{E}^u=\mathcal{E}_\beta $ for $\beta =s^{\prime }(u)$.

\item  \textit{Partial equivalence}. If $s$ is concave at $u$ but not
strictly concave, then $\mathcal{E}^u\subsetneq \mathcal{E}_\beta $ for
$\beta =s^{\prime }(u)$; i.e., $\mathcal{E}^u$ is a proper subset of
$\mathcal{E}_\beta $ which does not coincide with $\mathcal{E}_\beta $.
According to items 2 and 3, thermodynamic equivalence of ensembles for some
value of $u$ implies either full or partial equivalence of ensembles at the
level of equilibrium macrostates for that $u$.

\item  \textit{Nonequivalence.} If $s$ is nonconcave at $u$, then $\mathcal{E
}^u\cap \mathcal{E}_\beta =\emptyset $ for all $\beta $. In other words, if
there is thermodynamic nonequivalence of ensembles for some value of $u$,
then the microcanonical equilibrium macrostates corresponding to that $u$
are nowhere realized within the canonical ensemble.
\end{enumerate}

We spend the remainder of this section sketching part of the proof of full
equivalence as stated in item 2 and proving nonequivalence as stated in item
4. The crucial insight needed to prove that $s(u)=s^{**}(u)$ implies
$\mathcal{E}^u=\mathcal{E}_\beta $ with $\beta =s^{\prime }(u)$ is provided
by a basic result of convex analysis which states that 
\begin{equation}
s(v)\leq s(u)+\beta (v-u)
\end{equation}
for all $v$ if and only if $s(u)=s^{**}(u)$ and $\beta =s^{\prime }(u)$.
Accordingly, if we suppose that $s(u)=s^{**}(u)$, then 
\begin{equation}
\beta u-s(u)\leq \beta v-s(v)
\end{equation}
for all $v$. Using the thermodynamic representation of $\varphi (\beta )$ in
(\ref{lf1}) and the macrostate representation of $\varphi (\beta )$ in
(\ref{lf3}), we see that 
\begin{eqnarray}
\beta u-s(u) &\leq &\inf_{v\in \mathcal{U}}\{\beta v-s(v)\} \\
&=&\varphi (\beta )  \nonumber \\
&=&\inf_{L\in \mathcal{L}}\{\beta u(L)-s(L)\}.  \nonumber
\end{eqnarray}
Now choose any $L^u\in \mathcal{E}^u$; by the definition of this set
$u(L^u)=u$, and $s(L^u)=s(u)$. This allows us to write 
\begin{equation}
\beta u(L^u)-s(L^u)\leq \inf_{L\in \mathcal{L}}\{\beta u(L)-s(L)\}.
\end{equation}
We deduce that $L^u$ minimizes $\beta u(L)-s(L)$ or equivalently that
$L^u\in \mathcal{E}_\beta $. Since $L^u$ is an arbitrary element of
$\mathcal{E}^u$, it follows that $\mathcal{E}^u\subseteq \mathcal{E}_\beta $.
One completes the proof that $\mathcal{E}^u=\mathcal{E}_\beta $ by showing that
if $s$ is strictly concave at $u$, then $\mathcal{E}^u$ is not a proper
subset of $\mathcal{E}_\beta $.

To prove the assertion on nonequivalence of ensembles given in item 4, we
proceed in a similar fashion. The assumption that $s$ is nonconcave at $u$
implies that 
\begin{eqnarray}
s(u)
<
s^{**}(u) &=&\inf_\gamma \{\gamma u-\varphi (\gamma )\} \\
&\leq &\beta u-\varphi (\beta )  \nonumber
\end{eqnarray}
for all $\beta $. This can be rewritten as 
\begin{equation}
\beta u-s(u)>\varphi (\beta ).
\end{equation}
Now choose any $L^u\in \mathcal{E}^u$ and any $\beta $. Since $u(L^u)=u$ and 
$s(L^u)=s(u)$, it follows that 
\begin{equation}
\beta u(L^u)-s(L^u)>\varphi (\beta )=\inf_{L\in \mathcal{L}}\{\beta
u(L)-s(L)\}.
\end{equation}
This shows that $L^u$ is not a minimizer of $\beta u(L)-s(L)$ and thus that
$L^u\notin \mathcal{E}_\beta $. Since $L^u$ is an arbitrary element of
$\mathcal{E}^u$ and $\beta $ is arbitrary, we conclude that $\mathcal{E}
^u\cap \mathcal{E}_\beta =\emptyset $ for all $\beta $, as claimed.

\section{Illustration of the results for the Blume-Emery-Griffiths model}

\label{sBEG}

We now come to the main point of our study which is to illustrate, in the
context of a simple spin model, the general results presented in the
previous section about macrostate equivalence and nonequivalence of
ensembles and the relationship with thermodynamic equivalence and
nonequivalence. The model that we consider for this purpose is a slight
variant of the mean-field Blume-Emery-Griffiths (BEG) model defined by the
following Hamiltonian \cite{blume1971,barre2001}:
\begin{equation}
U(x^n)=\sum_{i=1}^nx_i^2-\frac{K}{n} \left( \sum_{i=1}^nx_i\right) ^2.
\end{equation}
In this formula $x_i$ represents a spin variable at site $i$ taking values
in the set $\mathcal{X}=\{-1,0,+1\}$, and $K$ is a positive real constant.
The a priori measure for the model is defined by
\begin{equation}
P(x^n)=\frac 1{3^n}\textrm{ for every }x^n\in \mathcal{X}^n.
\end{equation}

The macroscopic variable that we use to investigate the equivalence and
nonequivalence of the microcanonical and canonical ensembles is the
\textit{empirical vector} $L(x^n)$ defined as
\begin{equation}
L(x^n)=(L_{-1}(x^n),L_0(x^n),L_{+1}(x^n)),
\end{equation}
where for $j=-1,0,+1$
\begin{equation}
L_j(x^n)=\frac 1n\sum_{i=1}^n\delta _{x_i,j}.
\end{equation}
In physics, this quantity is often referred to as the \textit{one-particle
state distribution} or \textit{statistical distribution}; its three
components $L_{-1}(x^n)$, $L_0(x^n)$, and $L_{+1}(x^n)$ give the proportion
of spins in the microstate $x^n$ that take the respective values $-1$, $0$,
and $+1$. Accordingly, $L(x^n)$ takes values in the space $\mathcal{L}$ of
probability vectors $L=(L_{-1},L_0,L_{+1})$ defined by the conditions
$L_j\geq 0$ for $j=-1,0,+1$, and $L_{-1}+L_0+L_{+1}=1$.

For $L\in \mathcal{L}$ we define the function
\begin{equation}
u(L)=L_{+1}+L_{-1}-K(L_{+1}-L_{-1})^2.  \label{defineu}
\end{equation}
The choice of the empirical vector for studying equivalence and
nonequivalence of ensembles is dictated by the fact that we can represent
the energy per particle
\begin{equation}
u(x^n)=\frac{U(x^n)}n=\frac 1n\sum_{i=1}^nx_i^2-K\!\left( \frac
1n\sum_{i=1}^nx_i\right) ^2
\end{equation}
in the form
\begin{eqnarray}
u(x^n) &=&\sum_{j\in \mathcal{X}}L_j(x^n)j^2-K\!\left( \sum_{j\in \mathcal{X}
}L_j(x^n)j\right) ^2 \\
&=&L_{+1}(x^n)+L_{-1}(x^n)-K(L_{+1}(x^n)-L_{-1}(x^n))^2  \nonumber \\
&=&u(L(x^n)).  \nonumber
\end{eqnarray}
This display, which holds exactly for
all $x^n$ and $n$, verifies assumption (\ref{ulxn}) for the BEG model with
the energy representation function $u(L)$ defined in (\ref{defineu}).
Moreover, one shows either by a combinatorial argument based on Stirling's
approximation \cite{reif1965,pathria1996,ellis1985} or from Sanov's Theorem
\cite{dembo1998,ellis1985} that with respect to the a priori probability
measure $P$, $L(x^n)$ satisfies an LDP of the form (\ref{ldp2}) with the
entropy function
\begin{equation}
s(L)=-\sum_{x\in \mathcal{X}}L_x\ln L_x-\ln 3.
\end{equation}

These properties of $L(x^n)$ allow us to characterize the equilibrium
macrostates with respect to each ensemble as solutions of an appropriate
optimization problem. In order to simplify the notation, the components of
probability vectors $L\in \mathcal{L}$ will be written as $(L_{-},L_0,L_{+})$
instead of as $(L_{-1},L_0,L_{+1})$. We first consider the set $\mathcal{E}
^u $ of microcanonical equilibrium empirical vectors $L^u$ associated with
the mean energy $u$. According to the definition (\ref{micro1}), the
equilibrium macrostates $L^u$ are characterized as maximizers of $s(L)$ over
the space $\mathcal{L}$ subject to the constraint $u(L^u)=u$. Solving this
problem necessitates only the maximization of a function of one variable
since the normalization constraint on the components of the empirical vector
reduces the number of independent components of $L^u$ to two, while the
microcanonical energy constraint reduces this number by one more. The set
$\mathcal{E}_\beta $ of canonical equilibrium empirical vectors parameterized
by the inverse temperature $\beta $ is defined in (\ref{can1}). The elements 
$L_\beta $ of this set are characterized as maximizers of the quantity
$\beta u(L)-s(L)$ over $\mathcal{L}$. In this case, we are faced with an
unconstrained two-dimensional maximization problem involving the two
components $L_{+}$ and $L_{-}$.\footnote{
Another method for constructing 
$\protect\mathcal{E}_{\protect\beta}$\
can be based on the determination of the canonical equilibrium value of the
total spin per particle. The advantage of this alternate method is that the
associated minimization problem is one-dimensional (R.\ S.\ Ellis, P.\ Otto,
and H.\ Touchette, in preparation).}

\begin{figure*}
\resizebox{0.75\textwidth}{!}{\includegraphics{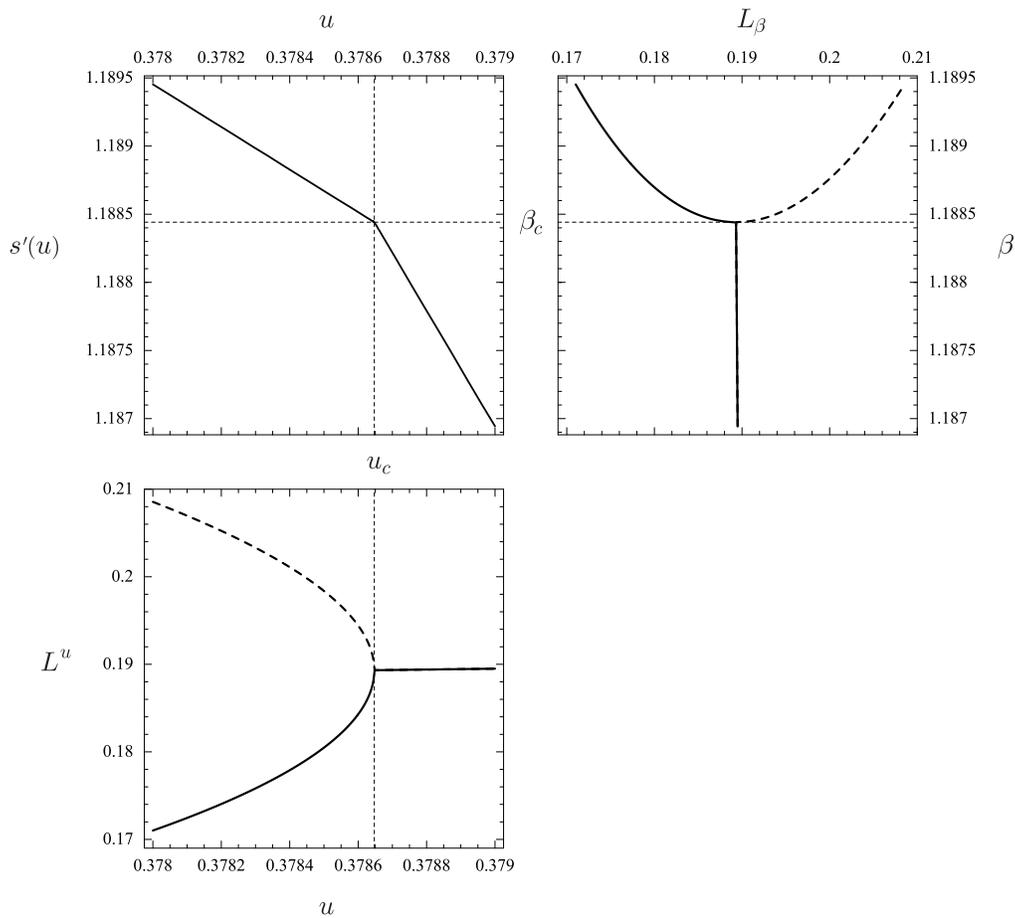}}
\caption{Full equivalence of ensembles
for the BEG model with $K=1.1111$.
(Top left) Derivative of the microcanonical entropy $s(u)$.
(Top right) The components $L_+$ and $L_-$ of the equilibrium
empirical measure $L_\beta$ in the canonical ensemble
as functions of $\protect\beta$.
For $\beta > \beta_c$ the solid and dashed
curves can be taken to represent $L_+$ and $L_-$, respectively, and vice versa.
(Bottom left) The components $L_+$ and $L_-$ of the equilibrium
empirical measure $L^u$ in the microcanonical ensemble
as functions of $u$. For $u < u_c$ the solid and
dashed curves can be taken to
represent $L_+$ and $L_-$, respectively, and vice versa.}
\label{k0}
\end{figure*}

In Figure \ref{k0} we present a first set of solutions for $\mathcal{E}^u$
and $\mathcal{E}_\beta $ corresponding to the value $K=1.1111$, together
with a plot of the derivative of the microcanonical entropy function $s(u)$.
Because neither of the two optimization problems involved in the definitions
of $\mathcal{E}^u$ and $\mathcal{E}_\beta $ could be solved analytically, we
provide numerical results obtained using various routines available in the
scientific software Mathematica. The top left plot of Figure \ref{k0}
showing $s^{\prime }(u)$ was obtained by calculating an empirical vector
$L^u\in \mathcal{E}^u$, which satisfies $u(L^u)=u$ and $s(L^u)=s(u)$. The top
right and the bottom left plots display, respectively, the canonical and
microcanonical equilibrium components of the empirical vector as a function
of the parameters $\beta $ and $u$ defining each of the two ensembles. In
the top right plot, the solid curve can be taken to represent the spin $+1$
component of the equilibrium empirical vector $L_\beta $, while the dashed
curve can be taken to represent the spin $-1$ component of the same
equilibrium empirical vector. Since the BEG\ Hamiltonian satisfies the
exchange symmetry $L_{+}\leftrightarrow L_{-}$, the roles of the solid and
dashed curves can also be reversed. For $\beta \leq \beta _c$, the solid
curve represents the common value of $L_{+}=L_{-}$. In all cases, the
component $L_0$ of $L_\beta $ is determined by $L_0=1-L_{+}-L_{-}$. The same
explanation applies to the bottom left plot of $L^u$.

The first series of plots displayed in Figure \ref{k0} were designed to
illustrate a case where $s(u)$ is concave and where, accordingly, we expect
equivalence of ensembles. That the equivalence of ensembles holds in this
case at the level of the empirical vector can be seen by noting that the
solid and dashed curves representing the $L_{+}$ and $L_{-}$ components of
$L_\beta $ in the top right plot can be put in one-to-one correspondence with
the solid and dashed curves representing the same two components of $L^u$ in
the bottom left plot. The one-to-one correspondence is defined by the
derivative of the microcanonical entropy $s(u)$: for a given $u$ we have
$L^u=L_{\beta (u)}$ with $\beta (u)=s^{\prime }(u)$. Moreover, since the
monotonic function $s^{\prime }(u)$ can be inverted to yield a function
$u(\beta )$ satisfying $s^{\prime}(u(\beta))=\beta$, we have $L_\beta
=L^{u(\beta )}$ for all $\beta $. Thus, the equilibrium statistics of the
BEG model in the microcanonical ensemble can be translated unambiguously
into equivalent equilibrium statistics in the canonical ensemble and vice
versa. In this case, the critical mean energy $u_c$ at which the BEG model
goes from a high-energy phase of zero magnetization $m(L)=L_{+}-L_{-}$ to a
low-energy phase of nonzero magnetization in the microcanonical ensemble can
be calculated from the viewpoint of the canonical ensemble by finding the
critical inverse temperature $\beta _c$ that determines the onset of the
same phase transition in the canonical ensemble. Since the two ensembles are
equivalent, both the microcanonical and canonical phase transitions must be
of the same order, which in this case is second-order.

\begin{figure*}
\resizebox{0.75\textwidth}{!}{\includegraphics{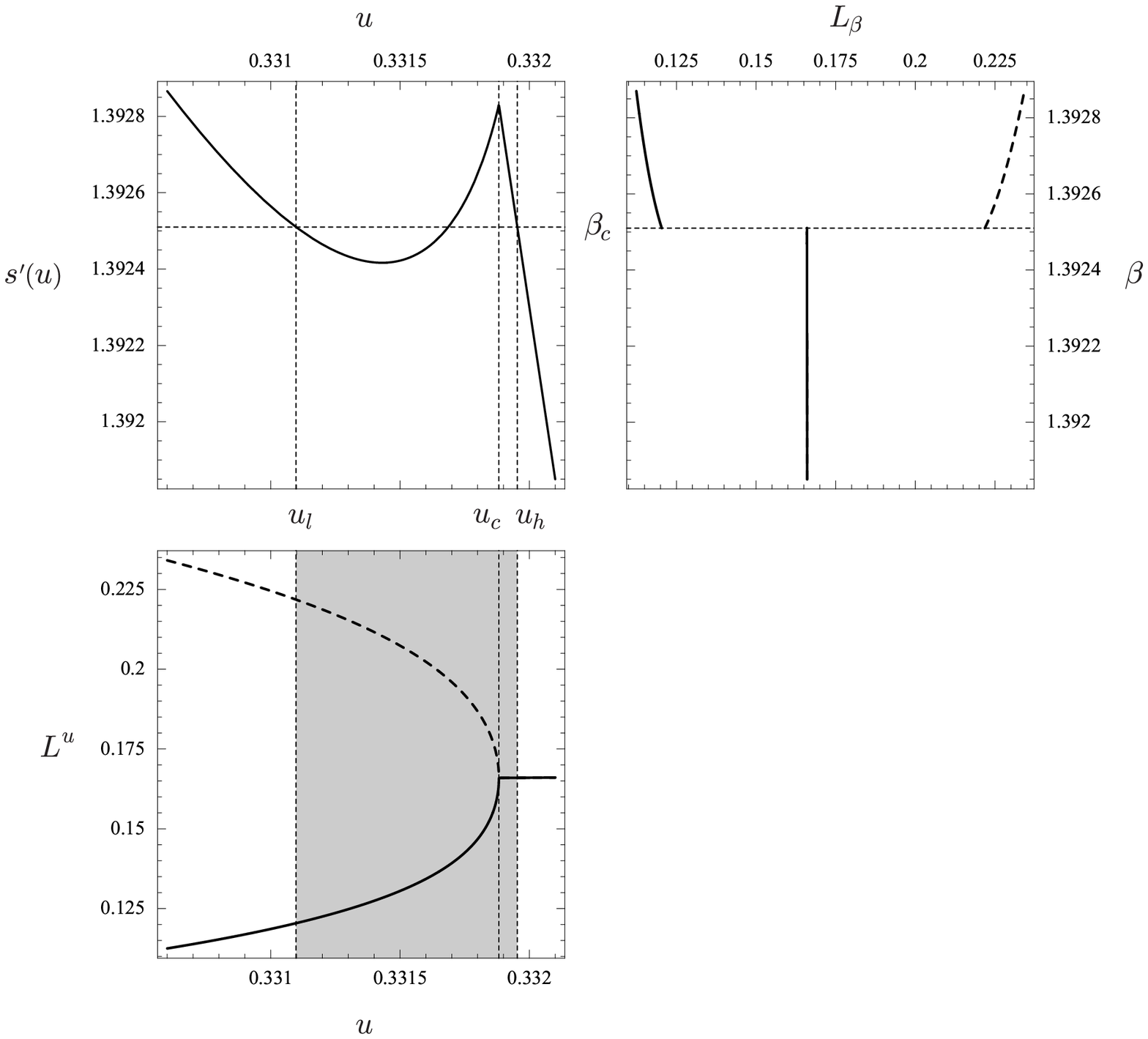}}
\caption{Equivalence and nonequivalence regions for the BEG
model with $K=1.0817$.  The solid and dashed curves
are interpreted as in Figure \protect\ref{k0}.  The shaded area
in the bottom left plot corresponds to the region of nonequivalence of
ensembles delimited by the open interval $(u_l,u_h)$. The ranges of
the inverse temperature and the mean energy used to draw the plots were chosen so as
to obtain a good view of the phase transitions.}
\label{k2}
\end{figure*}

In the second series of plots in Figure \ref{k2}, a case of ensemble
nonequivalence corresponding to the value $K=1.0817$ is shown. Since in the
top left plot $s^{\prime }(u)$ is not monotonic, $s(u)$ is not concave. As
in Figure \ref{fe3}, the open interval $(u_l,u_h)$ of mean-energy values is
the interval on which $s(u)\neq s^{**}(u)$; on this interval $s(u)$ is
nonconcave and $s^{**}(u)$ is affine with slope $\beta _c$. By comparing the
top right plot of $L_\beta $ and the bottom left plot of $L^u$, we see that
the elements of $\mathcal{E}^u$ cease to be related to elements of $\mathcal{
E}_\beta $ for all mean-energy values $u$ in the interval $(u_l,u_h)$. In
fact, for any $u$ in this interval of thermodynamic nonequivalence of
ensembles (shaded region) no $L_\beta $ exists that can be put in
correspondence with an equivalent equilibrium empirical vector contained in
$\mathcal{E}^u$. This lack of correspondence agrees with the rigorous results
of Ellis, Haven, and Turkington \cite{ellis2000} reviewed in Section
\ref{sEquivalence}. Thus, although the equilibrium macrostates $L^u$
corresponding to $u\in (u_l,u_h)$ are characterized by a well defined value
of the mean energy, it is impossible to assign a temperature to those
macrostates from the viewpoint of the canonical ensemble. In other words,
the canonical ensemble is blind to all mean-energy values $u$ contained in
the domain of nonconcavity of $s(u)$. By decreasing $\beta $ continuously
through the critical value $\beta _c$, the equilibrium value of the energy
per particle associated with the empirical vectors in $\mathcal{E}_\beta $
jumps discontinuously from $u_l$ to $u_h$ (canonical first-order phase
transition). However, outside the range $(u_l,u_h)$ we have equivalence of
ensembles, and a continuous variation of $\beta $ induces a continuous
variation of $u$.

We can go further in our analysis of the plots of Figure \ref{k2} by noting
that the phase transition exhibited in the microcanonical ensemble is
second-order (continuous) whereas it is first-order (discontinuous) in the
canonical ensemble. This provides another clear evidence of the
nonequivalence of the two ensembles. Again, because the canonical ensemble
is blind to all mean-energy values located in the nonequivalence region,
only a microcanonical analysis of the model can yield the critical mean
energy $u_c$. As for the critical inverse temperature $\beta _c$, which
signals the onset of the first-order transition in the canonical ensemble,
its precise value can be found by calculating the slope of the affine part
of $s^{**}(u)$ or, equivalently, by identifying the point of
nondifferentiability of $\varphi (\beta )$ (see the caption of Figure
\ref{fe3}).

A further characterization of $\beta _c$ can also be given in terms of the
three solutions of the equation $s^{\prime }(u)=\beta _c$. In Figure \ref{k2},
$u_l$ is the smallest of these solutions and $u_h$ the largest. We denote
by $u_m$ the intermediate solution of $s^{\prime }(u)=\beta _c$. Because
\begin{equation}
s(u_l)=s^{**}(u_l),\qquad s(u_h)=s^{**}(u_h),
\end{equation}
and
\begin{equation}
s^{**}(u_h)-s^{**}(u_l)=\beta _c(u_h-u_l),
\end{equation}
it follows that
\begin{equation}
\int_{u_l}^{u_h}[\beta _c-s^{\prime }(u)]\,du=\beta
_c(u_h-u_l)-[s(u_h)-s(u_l)]=0.
\end{equation}
Rewriting this integral in terms of $u_m$, we see that
\begin{equation}
\int_{u_l}^{u_m}[\beta _c-s^{\prime }(u)]\,du=\int_{u_m}^{u_h}[s^{\prime
}(u)-\beta _c]\,du.
\end{equation}
This equation expresses the equal-area property of $\beta _c$, first
observed by Maxwell (see \cite{huang1987}).

\begin{figure*}
\resizebox{0.75\textwidth}{!}{\includegraphics{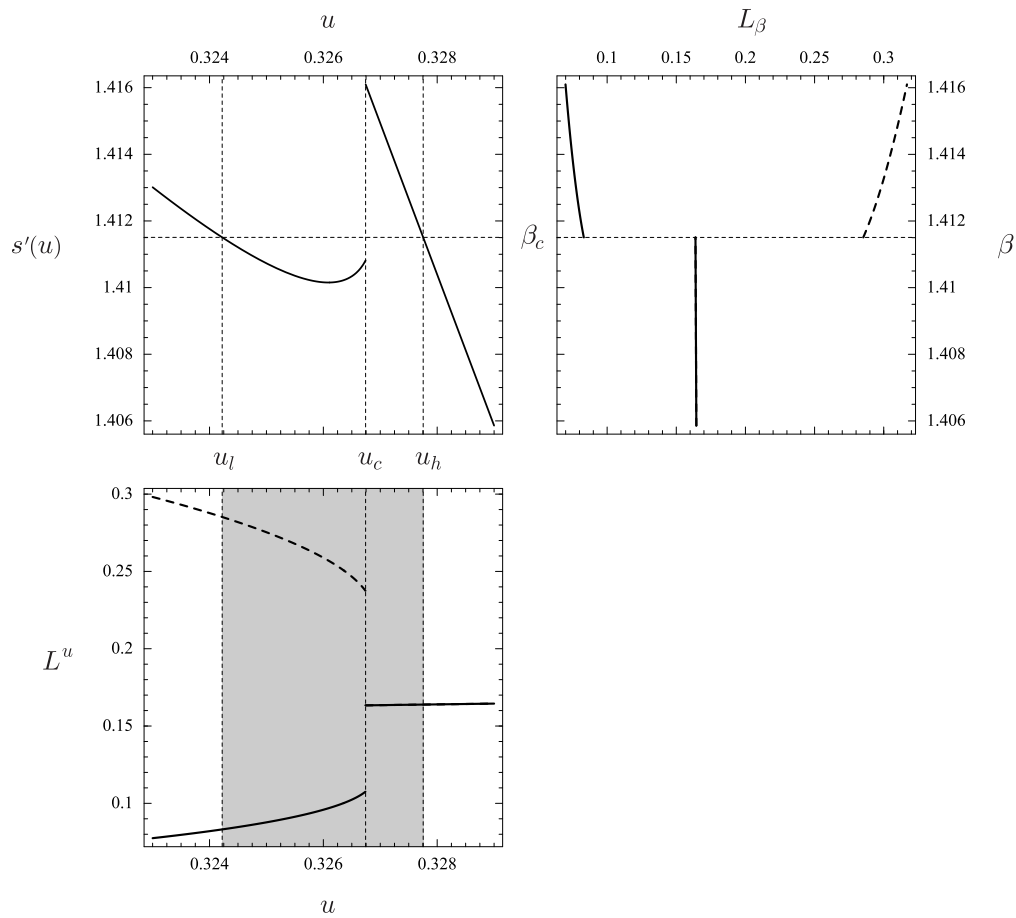}}
\caption{Equivalence and nonequivalence regions for the BEG
model with $K=1.0805$.  The solid and dashed curves
are interpreted as in Figure \protect\ref{k0}.  The shaded area
in the bottom left plot corresponds to the region of nonequivalence of
ensembles delimited by the open interval $(u_l,u_h)$.}
\label{k4}
\end{figure*}

To conclude this section, we present in Figure \ref{k4} a final series of
plots of $s^{\prime }(u)$, $L_\beta $, and $L^u$ corresponding to $K=1.0805$,
a slightly smaller value than the one considered in Figure \ref{k2}. As in
Figure \ref{k2}, there also exists in Figure \ref{k4} an open interval
$(u_l,u_h)$ over which $s(u)$ is nonconcave. For $u\in (u_l,u_h)$ we
consequently have nonequivalence of ensembles, illustrated by the shaded
region in the bottom left plot. As in Figure \ref{k2}, the nonequivalence of
ensembles is associated with a first-order phase transition in the canonical
ensemble determined by $\beta _c$. The microcanonical phase transition in
Figure \ref{k4} is also first-order due to the jump in $s^{\prime }(u)$ as
$u$ increases through the critical value $u_c$. By contrast, the
microcanonical transition is second-order in Figure \ref{k2}.

\section{Conclusion}

\label{sConclusion}

We have provided in this paper a simple illustration of the fact that the
nonequivalence of the microcanonical and canonical ensembles at the
thermodynamic level entails a much more fundamental nonequivalence of these
ensembles at the level of equilibrium macrostates. Focusing our attention on
the mean-field Blume-Emery-Griffiths (BEG) model, we showed that if the
microcanonical entropy $s$ is strictly concave at a mean-energy value $u$, then the
microcanonical equilibrium distributions of states characterizing the
equilibrium configurations of the BEG model at the fixed value $u$ are
realized canonically for inverse temperature $\beta $ given by $\beta
=s^{\prime }(u)$. We also showed that if the microcanonical entropy $s$ is
nonconcave at $u$, then the equilibrium distributions of states calculated
microcanonically at the fixed value $u$ are nowhere to be found in the
canonical ensemble.

This latter case of macrostate nonequivalence is illustrated in Figures \ref
{k2} and \ref{k4} for various values of the parameter $K$ entering in the
expression of the Hamiltonian of the BEG\ model. In each case there exists a
continuum of \textit{nonequivalent} equilibrium distributions of states that
can be associated with mean-energy values $u$ satisfying $s(u)<s^{**}(u)$,
but cannot be associated with values of the inverse temperature $\beta $,
the controllable parameter of the canonical ensemble. These results make it
clear that the microcanonical ensemble is richer than the canonical
ensemble, since the latter ensemble skips over the entire range of
mean-energy values for which the entropy is nonconcave. As we have seen,
this implies the presence of a first-order phase transition in the canonical
ensemble.

It should be remarked that although nonequivalent microcanonical equilibrium
distributions cannot be associated with any value of the parameter $\beta$,
one is not prevented from assigning to these distributions a
\textit{microcanonical} inverse temperature equal to the derivative of the 
microcanonical entropy. While this is a well-defined theoretical
possibility, important questions are whether this procedure has any
physical significance and whether such a microcanonical analog of inverse
temperature can be measured.

We end this paper by discussing another point related to nonequivalent
equilibrium macrostates corresponding to mean-energy values at which the
microcanonical entropy is nonconcave. In the course of doing our numerical
calculations leading to the determination of the sets $\mathcal{E}^u$ and
$\mathcal{E}_\beta $, we noticed that the equilibrium distributions of states 
$L^u$ that exist microcanonically but not canonically are \textit{metastable}
macrostates of the canonical ensemble; i.e., they are local but not global
minima of the quantity $\beta u(L)-s(L)$. This was observed to occur when
the mean energy $u$ associated with these states satisfies $s^{\prime \prime
}(u)<0$. For $u$ satisfying $s^{\prime \prime }(u)>0$, we found instead that
the macrostates $L^u$ in the region of nonequivalence are saddle points of
$\beta u(L)-s(L)$.

At this stage we cannot prove that this phenomenon holds in generality.
However, our results lead us to conjecture that it is valid for a wide
range of statistical mechanical models that have macroscopic variables
satisfying an LDP as in (\ref{ldp2}) with an entropy
function $s(L)$, and that have an energy representation function $u(L)$
satisfying (\ref{ulxn}). A number of theoretical and computational
results found by Eyink and Spohn \cite{eyink1993} and Antoni
et al.~\cite{antoni2002}, respectively, seem also to support 
such a conjecture. Besides, we know from the theory of Lagrange multipliers
that all the microcanonical equilibrium macrostates are extremal points
of the quantity $\beta u(L)-s(L)$, the global minimizers of which define
the canonical equilibrium macrostates. In this context, what remains to
be found is then just a way to determine the precise nature of these
extremal points based on the properties of $s(u)$. Work aimed at
elucidating this problem is ongoing \cite{touchette2003}.

\section*{Acknowledgments}

One of us (H.T.) would like to thank Stefano Ruffo for prompt and courteous
replies to many questions related to this work, and Andr\'{e}-Marie Tremblay
for useful discussions. Alessandro Torcini is also
thanked for pointing out Ref.~\cite{antoni2002} to our attention. The research of R.S.E.\ and
B.T.\ was supported by grants from the National Science Foundation
(NSF-DMS-0202309 and NSF-DMS-0207064, respectively) while the research of
H.T.\ was supported by the Natural Sciences and Engineering Research Council
of Canada (ESB Scolarships Program) and the Fonds qu\'{e}becois de la recherche
sur la nature et les technologies (Graduate Scolarships Program).

%\onecolumngrid
\bibliography{beglanl}

\end{document}